\documentclass{aastex}
\usepackage{emulateapj5}
\usepackage{apjfonts}

\def\wig#1{\mathrel{\hbox{\hbox to 0pt{%
          \lower.5ex\hbox{$\sim$}\hss}\raise.4ex\hbox{$#1$}}}}

\shorttitle{Atmospheric Dynamics and Hot Jupiter Spectra}
\shortauthors{Fortney et al.}

\newcommand{\mj}{$M_{\mathrm{J}}$}
\newcommand{\rj}{$R_{\mathrm{J}}$}

\newcommand{\T}{TrES-1}
\newcommand{\hd}{HD 209458b} 

\newcommand{\he}{HD 189733b}

\slugcomment{\bf}
\slugcomment{Accepted to the Astrophysical Journal}

\begin{document}

\title{The Influence of Atmospheric Dynamics\\on the Infrared Spectra and Light Curves of Hot Jupiters}

\author{J. J. Fortney\altaffilmark{1}$^,$\altaffilmark{2}$^,$\altaffilmark{3}, C. S. Cooper\altaffilmark{4}, A. P. Showman\altaffilmark{4}, M. S. Marley\altaffilmark{1}, R. S. Freedman\altaffilmark{1}$^,$\altaffilmark{3} }

\altaffiltext{1}{Space Science and Astrobiology Division, NASA Ames Research Center, MS 245-3, Moffett Field, CA 94035; jfortney@arc.nasa.gov, mark.s.marley@nasa.gov, freedman@darkstar.arc.nasa.gov}
\altaffiltext{2}{Spitzer Fellow}
\altaffiltext{3}{SETI Institute, 515 North Whisman Road, Mountain View, CA 94043}
\altaffiltext{4}{Lunar and Planetary Laboratory and Department of Planetary Sciences, University of Arizona, Tucson, AZ, 85721; curtis@lpl.arizona.edu, showman@lpl.arizona.edu}

\begin{abstract}

We explore the infrared spectrum of a three-dimensional dynamical model of
planet \hd~as a function of orbital phase.  The dynamical model predicts
day-side atmospheric pressure-temperature profiles that are much more
isothermal at pressures less than 1 bar than one-dimensional
radiative-convective models have found.  The resulting day-side thermal spectra are very similar to
a blackbody, and only weak water absorption features are seen at short
wavelengths.  The dayside emission is consequently in significantly
better agreement with ground-based and space-based secondary eclipse data than any previous models, which predict strong flux peaks and deep absorption features.  At other orbital phases, absorption due to carbon monoxide and methane
is also predicted.  We compute the spectra under two treatments of atmospheric chemistry:
one uses the predictions of equilibrium chemistry, and the other uses
non-equilibrium chemistry, which ties the timescales of methane and carbon
monoxide chemistry to dynamical timescales.  As a function of orbital phase,
we predict planet-to-star flux ratios for standard infrared bands and all
\emph{Spitzer Space Telescope} bands.  In \emph{Spitzer} bands, we predict 2-fold to 15-fold variation in planetary flux as a function of orbital phase with
equilibrium chemistry, and 2-fold to 4-fold variation with non-equilibrium
chemistry.  Variation is generally more pronounced in bands from 3-10 $\mu$m than at longer wavelengths.  The orbital phase of maximum thermal emission in infrared bands is
15--45 orbital degrees before the time of secondary eclipse.  Changes in
flux as a function of orbital phase for \hd~should be observable with
\emph{Spitzer}, given the previously acheived observational error bars.  

\end{abstract}



\section{Introduction}

Astronomers and planetary scientists are just beginning to understand the atmospheres of the short period giant planets known as ``hot Jupiters'' or ``Pegasi planets.''  A key subset of these planets are those that transit the disk of their parent stars, which make them well-suited for follow-up studies.  The most well studied of these transiting hot Jupiters is the first to be discovered, \hd~\citep{Charb00,Henry00}, a 0.69 \mj~planet that orbits its Sun-like parent star at a distance of 0.045 AU.

Currently, considerable work on hot Jupiters is occuring on both the observational and theoretical fronts.  In the past few years, several groups have computed dynamical
atmosphere models for \hd~in an effort to understand the structure,
winds, and temperature contrasts of the planet's atmosphere \citep{Showman02,Cho03,CS05,Burkert05,CS06}.  If the planet
has been tidally de-spun and become locked to its parent star, dynamical
models are surely needed to understand the extent to which absorbed stellar energy is
transported onto the planet's permanent night side.  With the launch of the
\emph{Spitzer Space Telescope}, there is now a platform that is well-suited for
observations of the thermal emission from hot Jupiter planets.

\emph{Spitzer} observations spanning the time of the planet's
secondary eclipse (when the planet passes behind its parent star) have been
published for \hd~at 24 $\mu$m \citep{Deming05b}, \T~at 4.5 and 8.0 $\mu$m \citep{Charb05}, and
\he~at 16$\mu$m \citep{Deming06}.  In all cases, the observed quantity is the
planet-to-star flux ratio in \emph{Spitzer} InfaRed Array Camera (IRAC),
InfraRed Spectrograph (IRS), or Multiband Imaging Photometer for
\emph{Spitzer} (MIPS) bands.  The dual \T~observations are especially interesting because they allow for a determination of the planet's mid-infrared spectral slope.  Several efforts have also been made from the
ground to observe the secondary eclipse of \hd.  Although no detections have
been made, some important, occasionallly overlooked upper limits at K and L
band have been obtained.  These include \citet{Rich03b} around 2.3 $\mu$m,
\citet{Snellen05} in K band, and \citet{Deming05c} at L band.  These ground-based 
observations constrain the flux emitted by the planet in spectral bands where water
vapor opacity is expected to be minimal; therefore, emitted flux should be
high.

This influx of data has spurred a new generation of radiative-convective equilibrium
models, whose resulting infrared spectra can be compared with data
\citep{Fortney05, Burrows05b, Seager05, Barman05, Fortney06, Burrows06}.  See \citet{Marley06} and \citet{Charb06} for reviews.  The majority of these models are one-dimensional. Authors weight the incident stellar flux by 1/4, to simulate planet-wide average conditions, or by 1/2, to simulate day-side average conditions (with a cold night side).  \citet{Barman05} have
investigated two-dimensional models with axial symmetry around the planet's
substellar-antistellar axis and computed infrared spectra as a function of orbital phase.
\citet{Iro05} have extended one-dimensional models by adding heat transport
due to a simple parametrization of winds to generate longitude-dependent
temperature maps, but they did not compute disk averaged spectra for these models.  Very recently \citet{Burrows06} have also investigated spectra and light curves of planets with various day-night effective temperature differences, assuming 1D profiles for each hemisphere.  These one- and two-dimensional radiative-convective equilibrium models have had some success in matching \emph{Spitzer} observations, but \citet{Seager05} and \citet{Deming05c} have shown that ground-based data for \hd~do not indicate promiment flux peaks at 2.3 and 3.8 $\mu$m, which solar composition models predict.

The various dynamical models for \hd~\citep{Showman02,Cho03,CS05,Burkert05,CS06} are quite varied in their
treatment of the planet's atmosphere.  We will not review them here, as
that is not the focus of this paper.  In general, temperature contrasts in the
visible atmosphere are expected to be somewhere between 300 and 1000 K.  What
has been somewhat lacking for these dynamical models are clear
observational signatures, which would in principle be testable with
\emph{Spitzer} or other telescopes.  The purpose of this paper is to remedy
that situation.  Here we generate infrared spectra and light curves as a
function of orbital phase for the \citet{CS06} (hereafter: CS06) dynamical simulation.
We present the first spectra generated for three-dimensional models of the
atmosphere of \hd.

\section{Methods}
Here we take the first step towards understanding the effects of atmospheric dynamics on the infrared spectra of hot Jupiters.  A consistent treatment of coupled atmospheric dynamics, non-equilibrium chemistry, and radiative transfer would be a considerable task.  In a coupled scheme, given a three-dimensional \emph{P--T} grid at a given time step, with corresponding chemical mixing ratios, the radiative transfer scheme would solve for the upward and downward fluxes in each layer.  These fluxes would be wavelength dependent and would differ from layer to layer.  The thermodynamical heating/cooling rate, which is the vertical divergence of the net flux, would then be calculated.  The dynamics scheme would then use this heating rate, together with the velocities and \emph{P--T} profiles in the grid at the previous time step, to a calculate the chemical abundances, velocities, and \emph{P--T} profiles on the grid at the new time step.  The process steps
forward in time, as the radiative transfer solver again
finds the new heating/cooling rates.  The emergent spectrum of the planet could be found at any stage.  In our work presented here, we performed a simplified calculation that contains some aspects of what will eventually be included in a fully consistent treatment.

\subsection{Dynamical Model}
Our input pressure-temperature (\emph{P--T}) map is from CS06.  As the
dynamical simulations are described in depth in \citet{CS05} and CS06, we will
only give an overview here.  The CS06 model employs the ARIES/GEOS Dynamical
Core, version 2 (AGDC2; \citealt{Suarez95}). The AGDC2 solves the primitive
equations of dynamical meteorology, which are the foundation of numerous
climate and numerical weather prediction models \citep{Holton92, Kalnay02}.
The primitive equations simplify the Navier-Stokes equations of fluid
mechanics by assuming hydrostatic balance of each vertical column of
atmosphere.  Forcing is due to incident flux from the parent star, through a Newtonian radiative process described in CS06.  The CS06 model is forced from the one-dimensional radiative-convective equilibrium profile of \citet{Iro05}, which assumes globally averaged planetary conditions.

\begin{figure}
\includegraphics[scale=1, angle=-90]{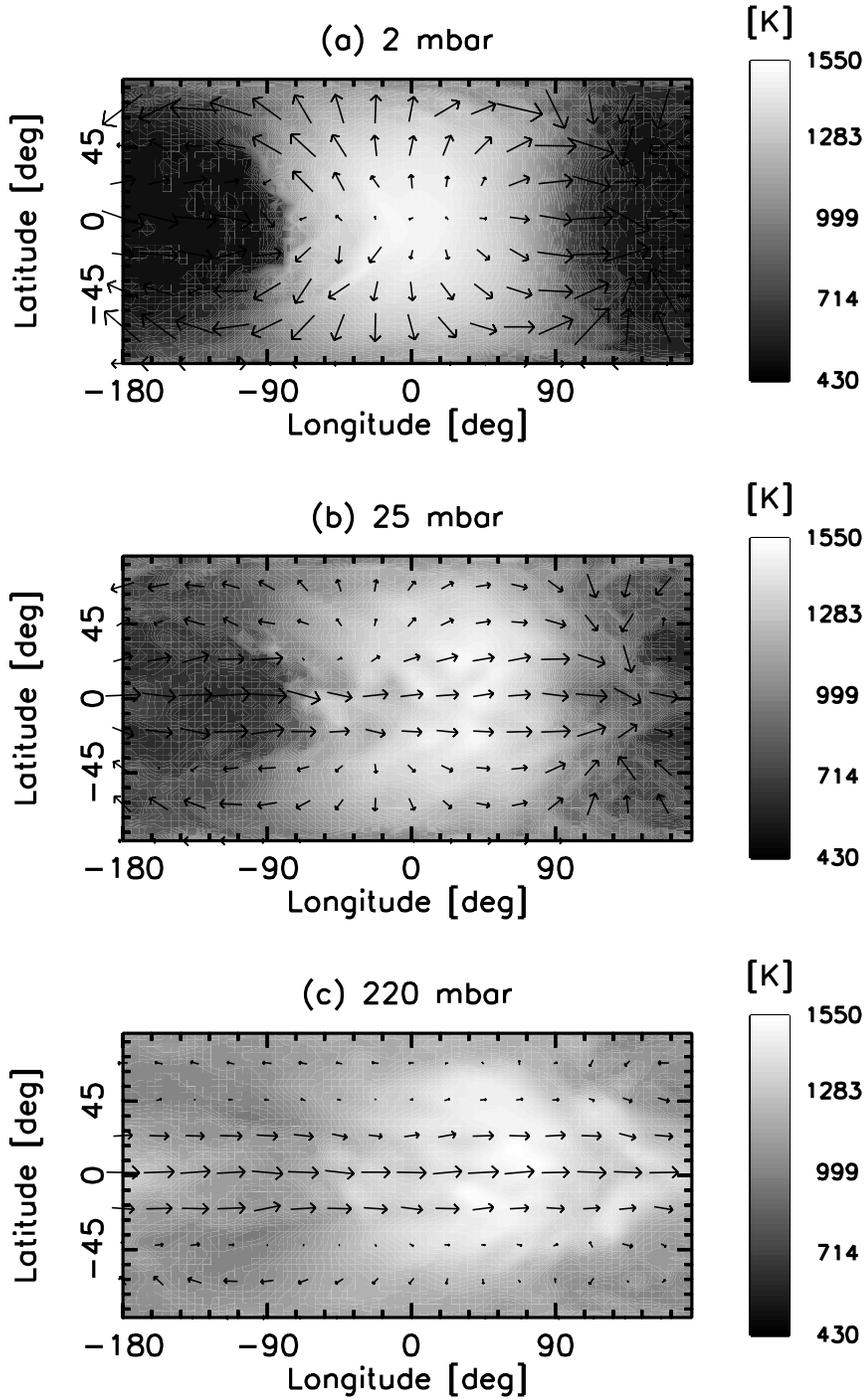}
\caption{Global temperature map at 2 mbar (top panel), 25 mbar (middle panel),
and 220 mbar (bottom panel) for the CS06 dynamical simulation.  Arrows show
the direction and relative magnitudes of winds.  Each longitude minor tick
mark is 18 degrees and each latitude minor tick mark is 9 degrees.  Each panel
uses the same temperature shading scheme.} 
\label{figure:dynam}
\end{figure}

For their simulations, CS06 take the top layer of the model to be 1 mbar.  The
model atmosphere spans $\sim$15 pressure scale heights between the input top
layer and the bottom boundary at 3 kbar.  CS06 use
40 layers evenly spaced in log pressure.  A \emph{P--T} profile is generated
at locations evenly spaced in longitude (in 5 degree increments) and latitude
(in 4 degree increments).  The 72 longitude and 44 latitude points create 3168
\emph{P--T} profiles.  In \mbox{Figure~\ref{figure:dynam}} we show the CS06
grid at 3 pressure levels, near $\sim$2, 20, and 200 mbar.  Previous work has
shown these levels likely bracket the pressures of interest for forming 
mid-infrared spectra \citep{Fortney05}.  All three panels of
\mbox{Figure~\ref{figure:dynam}} use the same brightness scale for easy
comparison between pressure levels.

At the 2 mbar level, where radiative time constants are short \citep{Iro05,
Seager05}, the atmosphere responds quickly to incident radiation.  Winds, though reaching a speed of up to 8 km s$^{-1}$, are not
fast enough to lead to significant deviations from a static atmosphere,
which implies that the hottest regions remain at the substellar
point.  The arrows indicate a wind pattern that attempts to carry
energy radially away from the hot spot.  At this pressure, the atmosphere
somewhat resembles the two-dimensional radially symmetric static atmosphere of
\citet{Barman05}, who found a hot spot at the substellar point and a uniformly
decreasing temperature as radial distance from this point increased.  The
night side appears nearly uniform, and colder.

At the 25 mbar level, it is clear that a west-to-east circulation pattern has
emerged at the equator, and the center of the planet's warm region has been
blown downstream by $\sim$35 degrees.  The wind from the west dominates over
the predominantly radially outward wind seen at the 2 mbar level.  Day-night
temperature contrasts are not as large at this pressure as they are at 2 mbar.

At the 220 mbar level, the center of the hot spot has been blown downstream by
$\sim$60 degrees.  This jet extends from the equator to the mid-latitudes; the
gas in the jet is warmer than gas to the north or south.  The radiative time
constants become longer the deeper one goes into the atmosphere \citep{Iro05}.
Hence, winds are better able to redistribute energy, leading to weaker
temperature contrasts, which cannot simply be characterized as ``day-night."

\subsection{Logistics \& Radiative Transfer}

Unlike other published models, we stress that the
spectra generated here are from a dynamical atmosphere model that is \emph{not} in
radiative-convective equilibrium.  Each of the 3168 \emph{P--T} profiles from CS06 have,
without modification (aside from interpolation onto a different pressure grid),
been run through our radiative transfer solver.  No iteration is done to
achieve radiative equilibrium.  The equation of radiative transfer is solved
with the two-stream source function technique described in \citet{Toon89}.
This is the same infrared radiative transfer scheme used in \citet{Fortney05}, \citet{Fortney06} and M.~S.~Marley et al.~(in prep.).  We ignore contributions due to scattered stellar
photons, as discussed below.

\begin{figure}
\plotone{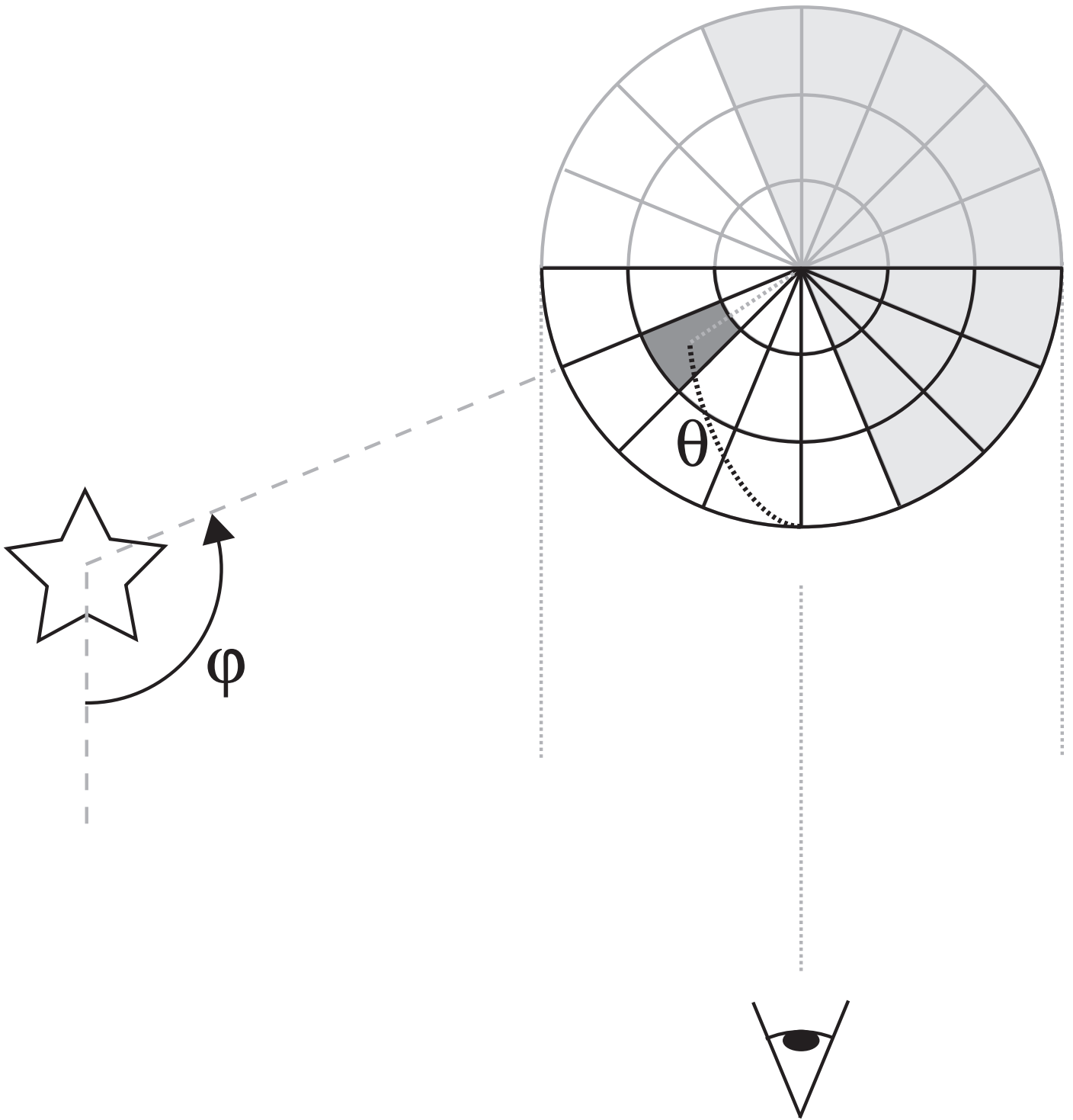}
\caption{Top-view diagram of the the planetary orbit.  The CS06
latitude/longitude grid covers the entire planet.  The planet's night side is
shaded gray.  At every angle in the orbit ($\varphi$), the longitude is
remapped to give the apparent longitude and latitude, as viewed by the distant
observer.  The cosine of the angle $\theta$ made between a patch of the planet
and the sub-observer point is $\mu$.
\label{figure:diagram}}
\end{figure}

At a given orbital position, the CS06 map in longitude is re-mapped into an
apparent longitude (as seen from Earth), while the latitude remains unchanged.
See \mbox{Figure~\ref{figure:diagram}} for a diagram.  Here we ignore the 3.4
degrees that the orbit differs from being exactly edge-on \citep{Brownetal01}.
At a given orbital angle $\varphi$---for each patch of the planet---the cosine of
the angle $\theta$ from the sub-observer point, $\mu$, is calculated.  
This $\mu$ is consistently included when solving the radiative transfer, which
means that the effects of limb darkening (or brightening) are automatically
incorporated.  We interpolate in a 
pressure-temperature-abundance grid \citep{Lodders02}, such that any given point in the three-dimensional model has the atomic and molecular abundances that are consistent with that point's pressure and temperature.

At any given time, the one-half of the planet that is not visible is not
included in the radiative transfer.  The 1584 visible points, at which the
emergent specific intensity (erg s$^{-1}$ cm$^{-2}$ Hz$^{-1}$ sr$^{-1}$) is
calculated, are then weighted by the apparent visible area of their respective
patches.  These intensities are summed up to give the total emergent flux
density (erg s$^{-1}$ cm$^{-2}$ Hz$^{-1}$) from the planet.  The spectra
generated by the patch-by-patch version of the code were tested against
spectra from one-dimensional gray atmospheres and our previously published hot
Jupiter profiles.

Emergent spectra are calculated from 0.26 to 325 $\mu$m, but since we ignore
the contribution due to scattered stellar flux, the spectra at the very
shortest wavelengths have little meaning.  However, for the
radiative-equilibrium \hd~model published in \citet{Fortney05}, they found that
scattered stellar flux is greater than thermal emission only at wavelengths
less than $\sim$0.68 $\mu$m.  We expect that a considerable amount of
``visible" light that may eventually be seen from hot Jupiters is due to
thermal emission.  We note that by 1 $\mu$m thermal emission is 100 times
greater than scattered flux.  Here we will present spectra for wavelengths
from 1 to 30 $\mu$m.

We note that the radiative transfer at every point is solved in the
plane-parallel approximation.  While this treatment is sufficient for our
purposes, we wish to point out two drawbacks.  The first is that, near the
limb of the planet, we will tend to overestimate the path lengths of photons
emerging from the atmosphere, as the curvature of the atmosphere is neglected.
The second issue also occurs near the limb.  We cannot treat photons whose
path, in a completely correct treatment, would start in one column but emerge
from an adjoining column.  However, atmospheric properties in any two
adjoining columns are in general quite similar.  The former issue, that of the
plane-parallel approximation, is likely more important, and should be
addressed at a later time, when data precision warrants it.  Here we note that
in our tests 95\% of planetary flux emerges from within 75 degrees of the
sub-observer point, such that these limb effects will have little effect on
the disk-summed spectra and light curves that we present.

When generating our model spectra, we use the elemental abundance data of
\citet{Lodders03} and chemical equilibrium compositions computed with the
CONDOR code, as described in \citet{Fegley94}, \citet{Lodders02}, and
\citet{Lodders02b}.  In \S2.3 we will discuss deviations from equilibrium
chemistry in the atmosphere of \hd, as calculated by CS06.  At this time, we
ignore photochemistry, which has been shown by \citet{Visscher06} to be
reasonable at $P>10$ mbar.  For the most part, the infrared spectra of hot
Jupiters are sensitive to opacity at $10 \lesssim P \lesssim 200$ mbar
\citep{Fortney05}, so equilibrium chemistry calculations are probably
sufficient.  As discussed in \citet{Fortney06}, we maintain a large and
constantly updated opacity database, which is described in detail in
R.~S.~Freedman \& K.~Lodders (in prep.).  The pressure, temperature,
composition, and wavelength-dependent opacity is tabulated beforehand using
the correlated-k method \citep{Goody89} in 196 wavelength interval bins.  The
resulting spectra are therefore of low resolution.  However, low resolution is
suitable for the task at hand, as we are interested in band-integrated fluxes
and the radiative transfer must be solved at 1584 locations on the planet at
many (here, 36) orbital phases.

In \citet{Williams06}, which focused on examining asymmetrical secondary
eclipse light curves caused by dynamical redistribution of flux, we
investigated the effects of limb darkening for \hd~for the CS06 map.  These
effects are hard to disentangle from general brightness variations due
to temperature differences generated by dynamics.  Limb darkening in a
particular wavelength would be manifested as a brightness temperature that
decreases towards the limb relative to a brightness temperature map computed
assuming normal incidence at every point.  On the planet's day side, where the
\emph{P--T} profiles are somewhat isothermal, limb darkening is not expected
to be significant.  Indeed, only at angles greater than $\sim$80 degrees from
the subsolar point was day-side limb darkening as large as 100--200 K
calculated.  As noted, due to our plane-parallel approximation, this is likely
to be somewhat of an overestimation.

\subsection{Clouds and Chemistry} 

\begin{figure}
\plotone{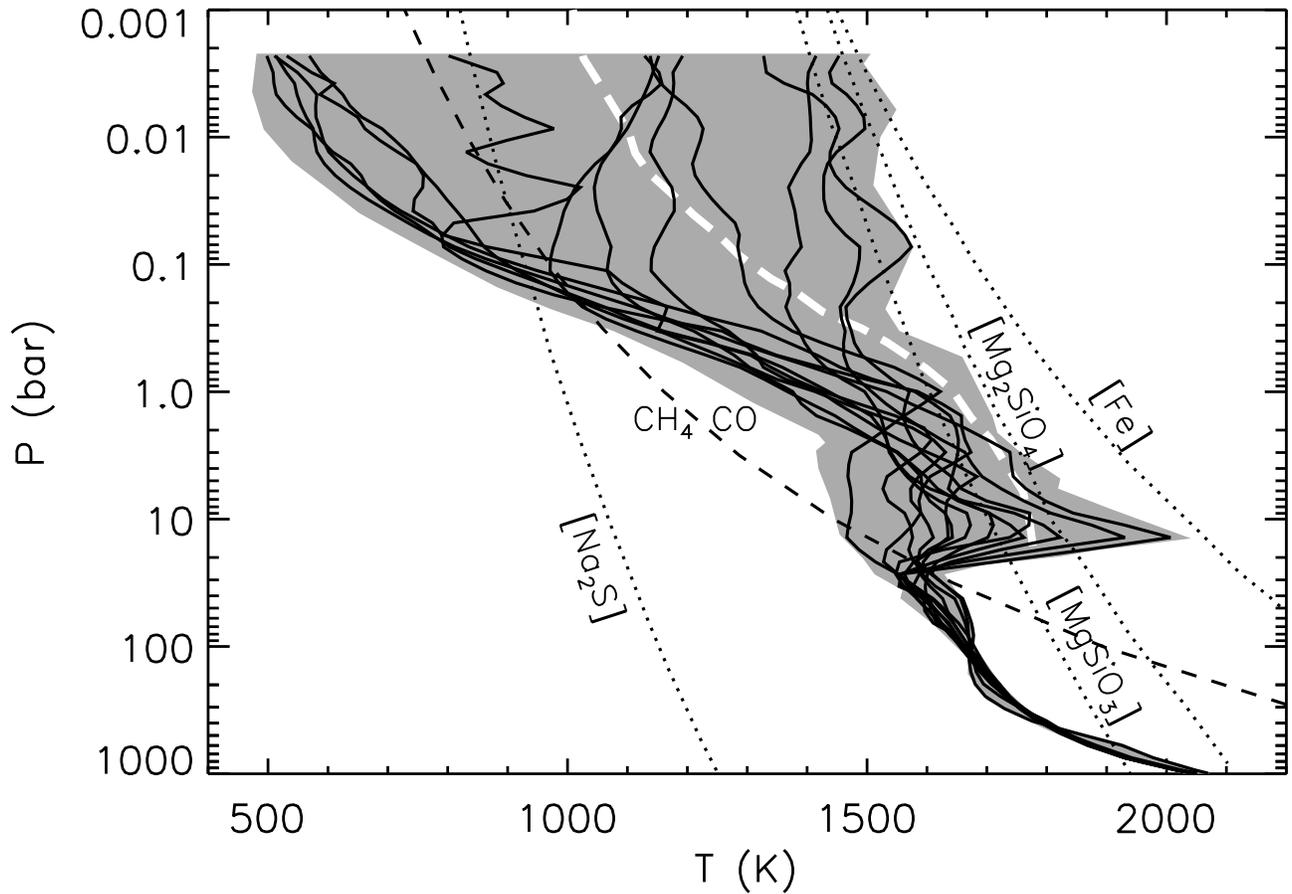}
\caption{In gray is the full pressure and temperature extent of the 3168
\emph{P--T} profiles from the CS06 simulation.  In thick black are 13 randomly
selected profiles that allow one to make out individual profile shapes and
temperatures.  In the middle of the gray, shown in white long dashes, is the
\citet{Iro05} profile.  The labeled dotted curves show the condensation curves
of important cloud species.  The dashed curve shows where CH$_4$ and CO have
an equal abundance.  All of these boundaries assume solar abundances.  Note
that day-side and limb profiles tend to be generally isothermal at $P < 200$
mbar.  The night-side profiles tend to plot over one another on the far
left.
\label{figure:PTs}}
\end{figure}

Before calculating spectra for this dynamic atmosphere, it is worthwhile to
step back and look at the atmospheric \emph{P--T} profiles with an eye towards
understanding the effects that clouds and chemistry may have on the emergent
spectra.  In \mbox{Figure~\ref{figure:PTs}}, we plot a random sampling of the
3168 \emph{P--T} profiles and compare them to cloud condensation and chemical
equilibrium boundaries.  It should be noted that there is a greater density of
profiles on the left side of the plot.  These are profiles from the relatively uniform and cool night
hemisphere.  Also of note is that at pressures less than $\sim$200 mbar, the warmer
profiles are fairly isothermal, with quite shallow temperature gradients.

The CS06 profiles are nearly everywhere cooler than the condensation curves of
iron and Mg-silicates.  These clouds will form, but for these profiles, cloud bases would lie deep in the atmosphere at pressures greater than 1 kbar, as
this is the highest pressure at which the profiles cross the condensation
curves of these species.  The opacity from such deep clouds would have \emph{no
effect} on the spectrum of \hd.

The temperatures of the CS06 simulation are computed as departures from the equilibrium temperature profile of \citet{Iro05}, as
described in CS06.  If a warmer base profile had been selected, such as
\citet{Fortney05} or \citet{Barman05} (which are 100-300 K warmer at our
pressures of interest), these curves could be shifted to the right by 100-300
degrees.  \citet{Marley06} provide a
graphical comparison of profiles computed by \citet{Iro05}, \citet{Fortney05},
and \citet{Barman05} under similar assumptions concerning redistribution of
stellar flux and atmospheric abundances; they find differences of up to 300 K
at 100 mbar.  These differences can probably be attributed to different
opacity databases, molecular abundances, radiative transfer methods, and perhaps incident stellar
fluxes.  For the \citet{Fortney05} profile, the major heating 
species are H$_2$O bands from 1 to 3 $\mu$m and neutral atomic Na and K, 
which absorb strongly in the optical.  The major cooling species are CO, 
at 5 
$\mu$m, and H$_2$O in bands at 3 and 4-10 $\mu$m, although for a colder 
profile, such as CS06 night side profiles, CH$_4$ would also cool across 
these 
wavelengths.  CS06 chose the 
\citet{Iro05} profile because Iro et al.~also
computed atmospheric radiative time constants, which other authors have not
done to date.  Here, the computed night side profiles cross the
condensation curve of Na$_2$S at low pressure.  \citet{Showman02} and
\citet{Iro05} pointed out that the transit detection of weak neutral atomic Na
absorption by \citet{Charb02} could be explained if a large fraction of Na was
tied up in this condensate.

If the predictions of equilibrium chemistry hold for this atmosphere, then it
would appear that at pressures up to 100 mbar, CH$_4$ would be the main carbon
carrier on the planet's night-side, while CO would be the main carrier on the
day side.  This would lead to dramatically different spectra for these
hemispheres, especially at near-infrared and mid-infrared wavelengths.

However, equilibrium chemical abundances would be expected only if chemical timescales are faster than any mixing timescales.  The relative abundances of CO and CH$_4$ can be driven out of equilibrium if mixing timescales due to dynamical winds are faster than the timescale for relaxation to equilibrium of the (net) reaction
\begin{equation}
\label{Cchem}
\mathrm{CO + 3H_2} \leftrightarrow \mathrm{CH_4 + H_2O}.
\end{equation}
The time constant for chemical relaxation toward equilibrium is a strong function of temperature and pressure; it
is short in the deep interior and extremely long in the observable atmosphere.  Deviations from CH$_4$/CO equilibrium are due to the long timescale for conversion of CO to CH$_4$, and are well known in the atmospheres of the Jovian planets \citep{Prinn77, Fegley85, Fegley94, Bezard02, Visscher05} and brown
dwarfs \citep{Fegley96, Noll97, Griffith99, Saumon00, Saumon03, Saumon06}, where vertical mixing can be due to  convection and/or eddy diffusion.  CS06 follow timescales for CH$_4$/CO chemistry taken from \citet{Yung88} and find that vertical winds of 5-10 m sec$^{-1}$ push this system out of equilibrium at pressures of interest for the formation of infrared spectra.  They find that the quench level, where the mixing timescale and chemistry timescale are equal, is $\sim$1-10 bar, which is below the visible atmosphere.  Above this quench level, the mole fractions of CO and CH$_4$ are essentially constant.  They find that the CH$_4$/CO ratio becomes homogenized at pressures less than 1 bar \emph{everywhere} on the planet.  For their nomical models this CH$_4$/CO ratio is $\sim$0.014, meaning that the majority of carbon is indeed in CO.  However, a non-negligible fraction of carbon remains in CH$_4$ on both the day and the night hemispheres.  We note that CS06 ignore possible effects due to photochemistry on the CO and CH$_4$ mixing ratios.  Given the few explorations into hot Jupiter carbon and oxygen photochemistry to date \citep{Liang03,Liang04,Visscher06}, it is unclear how important photochemistry will be in determining the abundances of these species at pressures of tens to hundreds of millibar.  CS06 did explore other effects such at atmospheric metallicity and temperature.  If the atmosphere is greater than [M/H]=0.0, or if the atmosphere is hotter, the CH$_4$/CO ratio would be even smaller.  See CS06 for additional discussion.

In our spectral calculations, we find that significant differences arise
depending on our treatment of chemistry.  We will therefore investigate the effects of a few chemistry cases, as explained below, and shown in Table 1.  We label our equilibrium
chemistry trial ``Case 0,'' as it is the standard case.  For ``Case 1,'' we fix the CH$_4$/CO ratio at 0.014 (as found by CS06) at all temperatures and pressures along the profiles.  Consistently incorporating the increasing CH$_4$/CO ratio at depth ($P>1$ bar) would be difficult with previously tabulated opacities,
and would have little to no effect on the emergent spectra.  Recall that 23\%
of available oxygen is lost to the formation of Mg-silicate clouds
\citep{Lodders03}, which in this model have cloud bases near 1 kbar.  The remaining
oxygen is almost entirely found in CO and H$_2$O.  As the CH$_4$ and CO abundances are fixed, for Case 1 we will also fix the abundance of H$_2$O; the mixing ratios
of CH$_4$, CO and H$_2$O are consistent with the amount of available oxygen at
T$\sim$ 1200 K and pressures of tens of millibars.  We also briefly consider a ``Case 2,'' in which the
CH$_4$/CO ratio has been further reduced, in an ad-hoc manner, which could be due to quenching of the abundances at a hotter temperature or the photochemical destruction of CH$_4$.  The CH$_4$/CO ratio is reduced by nearly a factor of 500, to $3 \times 10^{-5}$.  The mixing ratios of CO and H$_2$O are nearly the same as in Case 1, although the slightly increased CO abundance (due to the drop in the abundance of CH$_4$ and conservation of carbon atoms) uses up some oxygen at the expense of H$_2$O.  This case is valuable for comparison purposes because it shows the spectral effects of a negligible CH$_4$ abundance.  In all cases, the mixing ratios of all other chemical species are given by equilibrium values.  However, only CH$_4$, CO, and H$_2$O have a discernable impact on the spectra.

\begin{figure}
\plotone{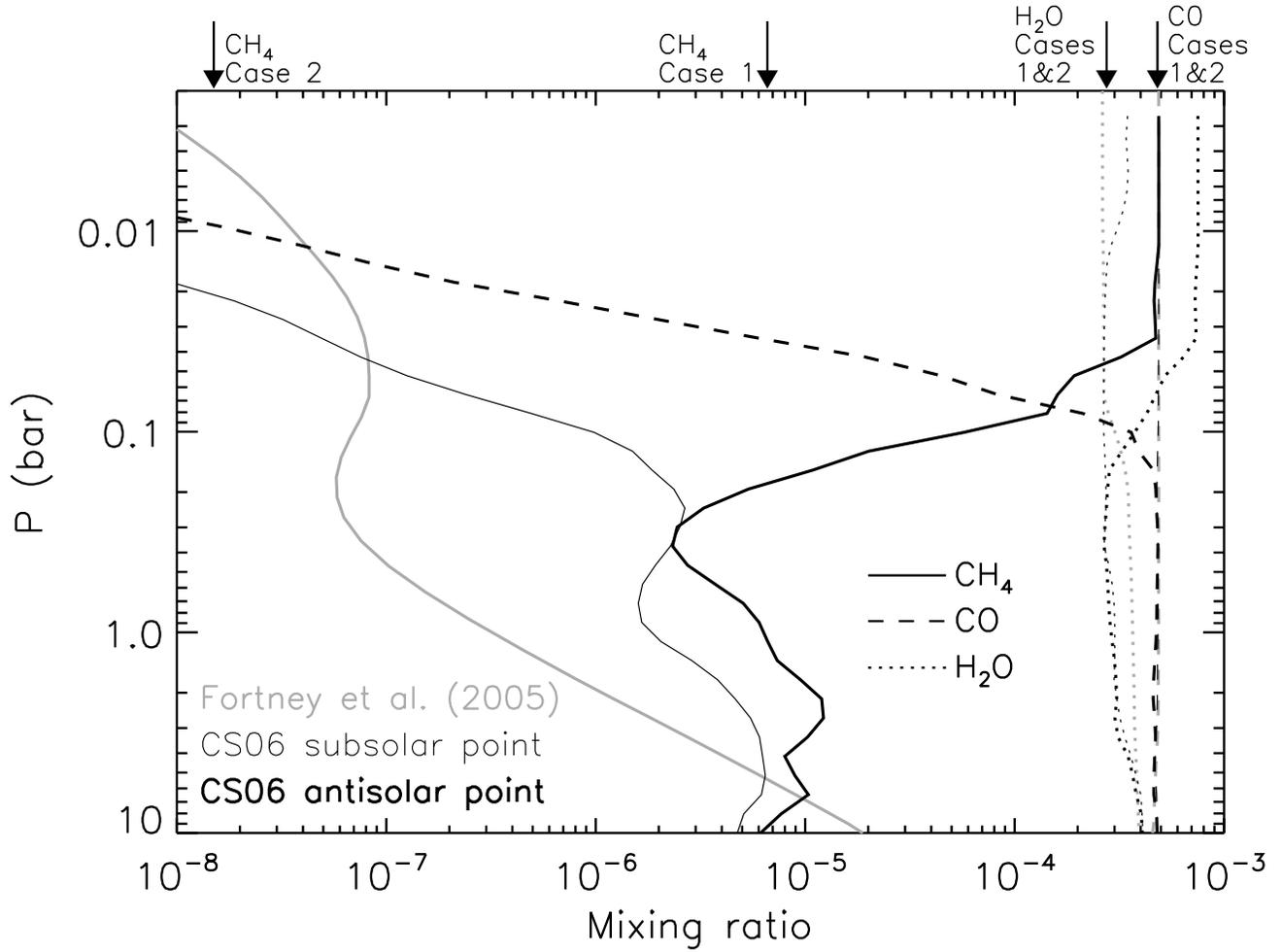}
\caption{The mixing ratios of CH$_4$, CO, and H$_2$O, as function of pressure, along three \emph{P--T} profiles in the atmosphere of \hd.  These profiles are the one-dimensional profile from \citet{Fortney05}, in gray, the subsolar point of the CS06 simulation, in thin black, and the anti-stellar point of the CS06 simulation, in thick black.  Arrows at the top of the plot indicate the fixed mixing ratios of Case 1 and 2 chemistry, given in Table 1. 
\label{figure:chem}}
\end{figure}

In \mbox{Figure~\ref{figure:chem}} we plot mixing ratios of CH$_4$, CO, and H$_2$O predicted from equilibrium chemistry along three \emph{P--T} profiles in the atmosphere of \hd.  In gray are the mixing ratios for the one-dimensional profile of \citet{Fortney05}.  The abundance of CH$_4$ is negligible.  In thin black and thick black are the mixing ratios at the sub-stellar point and the anti-stellar point, respectively, of the CS06 simulation.  The dominant carbon carrier is clearly CO, rather than CH$_4$, except at $P<$100 mbar for the anti-stellar point.  Most of the nightside has a chemistry profile similar to the anti-stellar point, and hence, strong CH$_4$ absorption will be seen.  At the top of the plot, arrows indicate the assigned mixing ratios of Cases 1 and 2 for these three molecules.

In \S3 we make quantitative comparisons to current ground-based and space-based infrared data.  While we find that the agreement is good for the model presented here, we wish to stress that this study is exploratory.  This is the first study that quantitatively explores the influence of atmospheric dynamics on the emergent spectra of a hot Jupiter atmosphere.  The greatest uncertainty likely lies in the calculation of heating/cooling rates with a simple Newtonian cooling scheme in the dynamical model, as discussed in \citet{CS05} and CS06.  Another issue is that CS06 calculate temperature deviations from a one dimensional radiative-convective profile published in \citet{Iro05}.  However, our emergent spectrum is calculated using our radiative transfer solver, which is a different code, and we use different abundances \citep{Lodders03} than these authors \citep{Anders89}.  However, we find that when solving the radiative transfer for the \citeauthor{Iro05} profile we obtain a $T_{\rm eff}$ that differs by only 1\% from their value.

In their non-equilibrium chemistry work, CS06 chose the abundances of \citet{Lodders98}, while here we use \citep{Lodders03}, in order for the most natural comparison with our previous work \citep{Fortney05,Fortney06}.  Obviously no choice would be fully self consistent.  However, the choice of elemental abundances is certainly a smaller concern than the current debate concerning the correct chemical timescale for conversion of CO to CH$_4$ in planetary atmospheres \citep[see][]{Yung88,Fegley94,Griffith99,Bezard02,Visscher05}.  In addion, CS06 have shown that a 300 K decrease in temperature, for instance, leads to a 20-fold increase in CH$_4$/CO (see \S5.3), so temperature uncertainties likely swamp any abundance issues.  We stress that our focus in the following sections is on highlighting the spectral differences between a model that accounts for dynamical redistribution of energy around the planet and one that does not.  Indeed the magnitude of the spectral effect suggests that additional, more internally self consistent work, is clearly appropriate. 

\section{Infrared Spectra}
\subsection{Spectra as a Function of Orbital Phase}

\begin{figure}
\includegraphics[scale=.7, angle=-90]{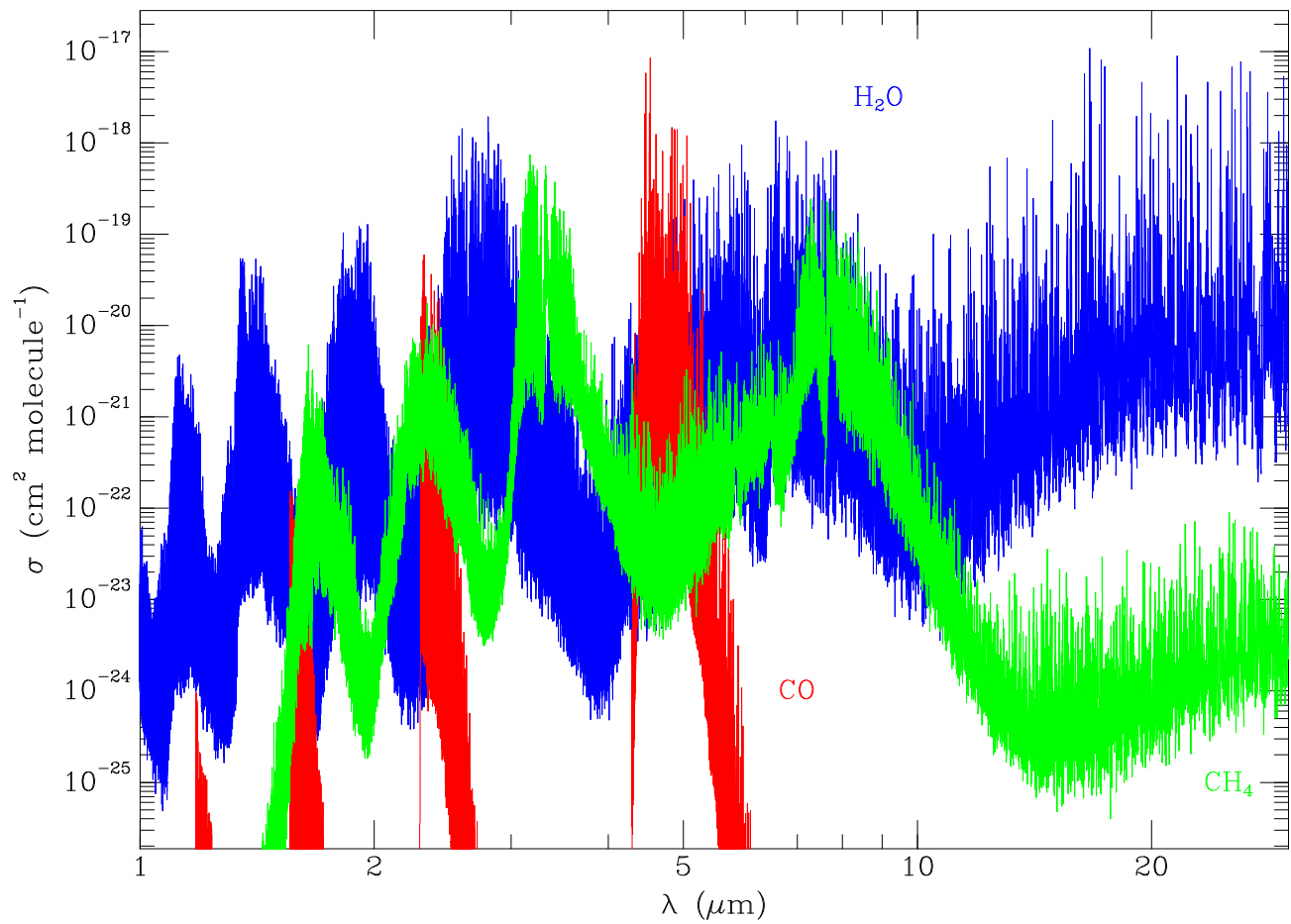}
\caption{Absorption cross-sections vs.~wavelength for H$_2$O (blue), CO (red), and CH$_4$ (green) at 1200 K and 100 mbar.  These cross-sections are not weighted by abundance.}
\label{figure:xsect}
\end{figure}

We now turn to the predicted infrared spectrum of our dynamical model
atmosphere.  As has been shown by many authors since \citet{SS98}, the infrared
spectra of hot Jupiters are believed to be carved predominantly by absorption
by H$_2$O, CO, and, if temperatures are cool enough, CH$_4$.  In general,
absorption features of hot Jupiters are predicted to be shallower than brown
dwarfs of similar effective temperature and abundances.  This is predicated on
hot Jupiters having a significantly shallower atmospheric temperature
gradient, which is due to the intense external irradiation by the parent star.
\mbox{Figure~\ref{figure:xsect}} shows absorption cross sections per molecule
for CH$_4$, CO, and H$_2$O from 1 to 30 $\mu$m.  To avoid clutter on our
spectral plots, we will not label absorption features, so referring to
\mbox{Figure~\ref{figure:xsect}} will be helpful.

\begin{figure}
\plotone{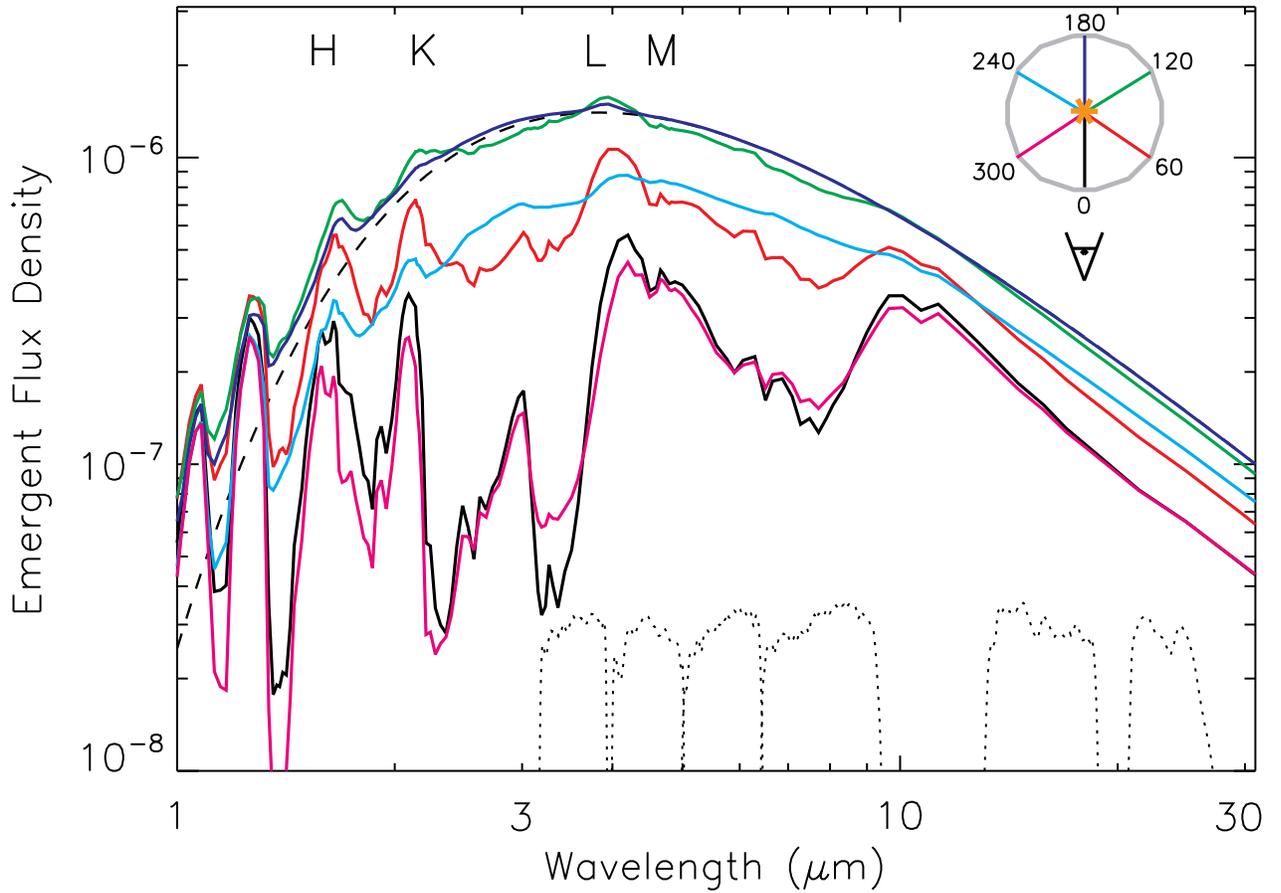}
\caption{Planetary emergent flux density (erg s$^{-1}$ cm$^{-2}$ Hz$^{-1}$)
vs.~wavelength as a function of orbital phase for Case 0 equilibrium
chemistry.  The spectra are color-coded with the diagram in the upper right
corner of the figure.  All spectra are calculated at intervals of 60 orbital
degrees.  The black spectrum is for the night side of the planet, which is
seen during transit.  The red spectrum is 60 orbital degrees later.  The dark
blue spectrum shows the planet during secondary eclipse (180 degrees), when
the fully illuminated hemisphere is visible.  The magenta spectrum is 300
orbital degrees after transit.  The dashed black curve is the flux of a 1330 K
blackbody, which plots behind the dark blue curve at $\lambda > 4 \mu$m.
Normalized \emph{Spitzer} band passes are shown in dotted lines at the bottom
and standard H, K, L, and M bands are shown at the top. 
\label{figure:6spec}}
\end{figure}

Our first spectral calculation is for Case 0 chemistry, as a function of
orbital phase.  This is shown in \mbox{Figure~\ref{figure:6spec}}.  We show
the emitted spectrum at six orbital locations: zero (during the transit), 60,
120, 180 (during the secondary eclipse), 240, and 300 degrees, where the
degrees are orbital degrees after transit.  At zero degrees (the black
spectrum), we see the full night side of the planet.  The temperature gradient
is fairly steep, leading to deep absorption features.  Most carbon is in the form
of CH$_4$, leading to strong methane absorption in bands centered on 1.6, 2.2,
3.3, and 7.8 $\mu$m.  Absorption due to H$_2$O at many wavelengths is also
strong.  Absorption due to CO at 2.3 and 4.5 $\mu$m is quite weak, due to its
small mixing ratio.

Sixty degrees later (the red spectrum), the hemisphere we see is 2/3 night and
1/3 day.  The part of the day hemisphere we see has the hot west-to-east jet
coming towards us.  This will later be an interesting point of comparison with
the 300 orbital degree spectrum (in magenta), which is also 2/3 night.  Here
all absorption features are muted because, due to the jet shown in
\mbox{Figure~\ref{figure:dynam}}, more than $\sim$1/3 of the planet is showing
the nearly isothermal profiles representative of the day side.  These profiles
lead to a blackbody-like spectrum.  At 120 orbital degrees (in green), we are
seeing 2/3 day side, and that part of the day side is the hottest, due to the
wind from the west blowing the hot spot downstream.  The absorption features
have become very weak. 

The peak in total infrared flux actually occurs 153 orbital degrees after transit.  During the secondary eclipse, 27 degrees later, the planet is less luminous.
During the secondary eclipse (in dark blue) the planetary spectrum is
essentially featureless long-ward of 1.8 $\mu$m.  The fully illuminated
hemisphere possesses profiles that are nearly isothermal in our pressure range
of interest.  Intriguingly, this leads to a spectrum that is very similar to a
blackbody.  For comparison, we plot a dashed black curve that is the thermal
emission of a 1330 K blackbody.  It plots behind the 180 degree spectrum for
all $\lambda > 4$ $\mu$m.

As we move to orbital phases after the secondary eclipse, we see that the
spectra are not symmetric with their reflections on the other half of the
orbit.  The light blue spectrum is significantly lower than the green one,
even though at both points we are seeing 2/3 day and 1/3 night.  This is again
due to the west-to-east jet.  At 240 orbital degrees, the hottest part of the
day side is not in view.  At 300 orbital degrees (in magenta) 1/3 of our view
is the coolest part of the day side, along with 2/3 of the night side.  The
spectrum is quite similar to what was seen during the transit.  The planet
appears least luminous 31 orbital degrees before transit.

\subsection{Spectra with Alternate Chemistry}

\begin{figure}
\plotone{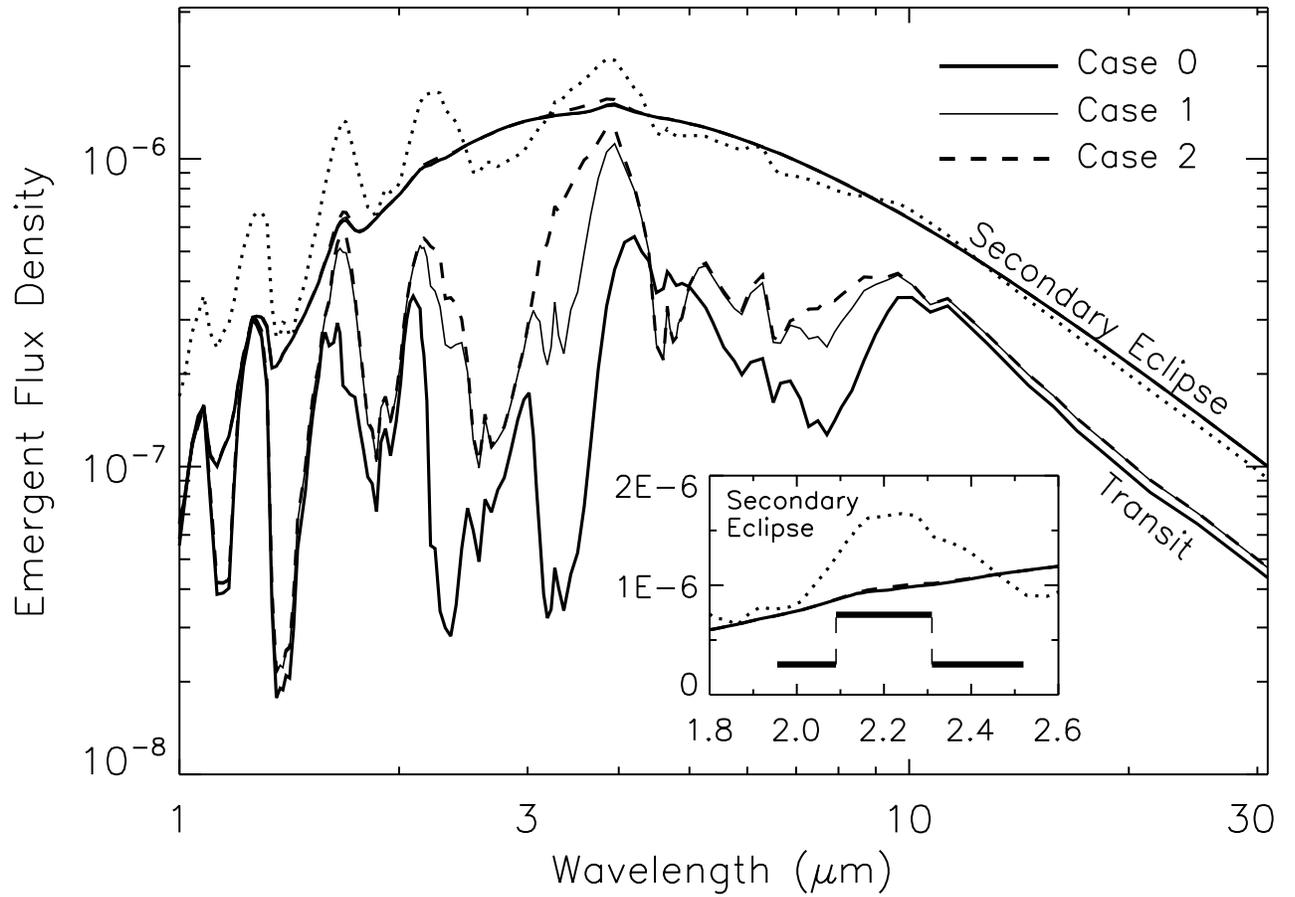}
\caption{Planetary emergent flux density (erg s$^{-1}$ cm$^{-2}$ Hz$^{-1}$)
vs.~wavelength at two orbital phases for all three of our chemistry cases.
The spectrum at 180 degrees, during the secondary eclipse, is essentially the
same for all three cases.  At zero degrees, during the transit, CH$_4$ is very
abundant in the Case 0 trial, leading to deep absorption features centered at
1.6, 2.2, 3.3, and 7.8 $\mu$m.  Absorption due to CO is muted in Case 0, but
is strong in Cases 1 and 2.  Water vapor absorption is strong in all cases.
\emph{Inset}:  Zoom near 2.2 $\mu$m.  The three thick horizontal lines
connected by dashes are a constraint on the maximum relative height of the
flux peak at 2.2 $\mu$m (middle band), compared to the average of the flux in
the two adjacent bands, from \citet{Seager05}.  The bars can be shifted
together vertically. (See text.)
\label{figure:cases}}
\end{figure}

Since CS06 find significant deviations from equilibrium carbon chemistry, it is useful to examine spectra with non-equilibrium chemistry.  We now investigate the spectra using Cases 1 and 2.
\mbox{Figure~\ref{figure:cases}} shows comparison spectra at two orbital
phases: zero and 180 orbital degrees, which is during transit and secondary
eclipse, respectively.  On the day side, we see that the emergent spectrum is
essentially exactly the same, independent of chemistry.  Because the
\emph{P--T} profiles are nearly isothermal, essentially no information about
the chemistry can be gleaned from the spectrum.  For comparison with our
previous work, in \mbox{Figure~\ref{figure:cases}}, we plot with a dotted line
the cloudless planet-wide average spectrum of \hd~from \citet{Fortney05},
which shows absorption due to H$_2$O and CO and strong flux peaks between
these absorption features.

We will now examine the small inset in \mbox{Figure~\ref{figure:cases}}.  This
is a zoom-in of the secondary-eclipse spectra near 2.2 $\mu$m.  The three
thick horizontal bars connected by dashes make up an interesting observational
one-sigma upper limit that was first reported in \citet{Rich03b}.  These data
were analyzed again and presented in \citet{Seager05}.  The measurement was a
relative one; only the vertical distance between the central band and two side
bands is important: they can be moved up or down as a group.  Specifically,
the flux in the central band cannot exceed the combined flux in the two
adjacent bands by an amount greater than vertical distance shown (which is
$0.45 \times 10^{-6}$ erg s$^{-1}$ cm$^{-2}$ Hz$^{-1}$).  Our dynamical day-side spectra, which are nearly
featureless due to the nearly isothermal atmosphere, easily meet this constraint.  We note that an upper limit
obtained by \citet{Snellen05} in K band, which covers a similar wavelength
range, has an error bar that is too large to distinguish among published models.

In contrast to the models presented here that include
dynamics, published 1D models to date all predict large flux differences between
2.2 $\mu$m and the surrounding spectral regions when
abundances are solar.   The 1D models with solar abundances
presented in \citet{Seager05}, for example, were not able to
meet the 2.2-$\mu$m flux-difference constraint, because the H$_2$O bands
adjoining the 2.2-$\mu$m flux peak were too deep.   \citet{Seager05} suggested that if C/O $>$ 1, then little H$_2$O would
exist in the atmosphere, and the observational constraint could be met.
Here, we propose instead that dynamics produces a relatively
isothermal dayside temperature, in which case the constraint
can easily be met even with solar C/O.

Nevertheless, the observational constraint is not yet firm enough
to fully rule out the 1D radiative-equilibrium models at solar
abundances.   The \citet{Fortney05} solar-abundance models,
for example, marginally meet the current observational flux-difference
constraint. (Fortney et al.~2005 find a flux difference of
$0.43 \times 10^{-6}$ erg s$^{-1}$ cm$^{-2}$ Hz$^{-1}$.)   The cause for the discrepancy
between the 1D models in \citet{Fortney05} and \citet{Seager05} remains unclear but could occur if the \emph{P--T} profiles
in Seager et al.~were steeper than those in Fortney et al.,
leading to deeper bands in the former study.  More precise
flux-difference observations like that of \citet{Rich03b} during secondary eclipse would help confirm or rule out
the 1D solar-abundance models.   Even a factor
of two decrease in the observational upper limit between the
``in'' and ``out'' bands could rule out the solar-abundance
\citet{Fortney05} model, while supporting the isothermal-dayside
models (e.g., CS06) and the C/O $>$ 1 models with weak water
bands.   Alternatively,  the detection of a small flux peak at
2.2 $\mu$m would lend support to the 1D models with C/O $<$ 1,
while constraining the temperature gradient on the planet's dayside.

The fact that both the solar-abundance CS06 models (presented here)
and the radiative-equilibrium 1D models with C/O $>$ 1 produce
almost no flux difference between 2.2 $\mu$m and the surrounding
continuum implies that the 2.2-$\mu$m band cannot be used to
distinguish between these alternatives.  Instead, additional constraints
at other wavelengths will  be necessary to discriminate between them.  Measurements of flux differences surrounding CO (rather than H$_2$O)
bands could provide such a test.  The radiative-equilibrium C/O $>$ 1 models,
although lacking water bands, would presumably have strong CO bands
and would hence predict a strong flux difference between the center of
a CO band and the surrounding continuum.  On the other hand,
because the dayside atmosphere is nearly isothermal in the CS06 models,
these circulation-altered models would predict little flux difference
between the CO band and the surrounding regions even in the presence
of large quantities of CO.   An observation of minimal flux difference
across CO bands as well as H$_2$O bands would support isothermal-dayside
models like CS06 while arguing against the radiative-equilibrium
1D models with steep temperature gradients. 

Although observations of \T~may not be directly applicable to \hd, as \T~may have a $T_{\rm eff}$ 300 K cooler, there is an important issue to note for our discussion here.  The models of \citet{Fortney05}, \citet{Seager05}, and \citet{Barman05} show a mid-infrared spectral slope that is bluer than observed from 4.5 to 8.0 $\mu$m by \citet{Charb05} for \T, although the model of \citet{Burrows05b} appears to be consistent with this slope.  \citet{Showman06} and \citet{Fortney06} have discussed that models with an atmospheric temperature inversion would give infrared spectra with a redder spectral slope, due to molecular emission features, in better agreement with observations.  The CS06 model of the day side of \hd, which lacks the strong negative temperature gradient of these radiative equilibrium models, also leads to a redder mid-infrared spectral slope.

On the planet's night side, we see significant spectral differences between
the three chemistry cases.  In Cases 1 and 2, the CH$_4$ mixing ratio is
constrained to a small abundance, weakening these absorption features.
Absorption due to CH$_4$, CO, and H$_2$O is readily seen in Case 1.  Case 2
shows essentially the same CO and H$_2$O absorption, but CH$_4$ absorption is
no longer seen.  The transition, as a function of orbital phase from deep to
essentially nonexistent absorption features in Cases 1 and 2, are similar to
what was seen in Case 0.  In the interest of conciseness, and because Case 2
is ad-hoc, hereafter we concentrate on Cases 0 and 1.  The spectra for
Cases 1 and 2 are essentially the same, except in regions of CH$_4$
absorption; we will highlight these differences when necessary.  It is
important to remember that in Case 1 and Case 2 chemistry, the mixing
ratios of our principal absorbers, CH$_4$, CO, and H$_2$O are fixed.
Therefore, changes in the spectra with orbital phase are only due to changes in the planetary \emph{P--T} profiles on the visible disk, due to the rotation of the planet
through its orbit.

\begin{figure}
\plotone{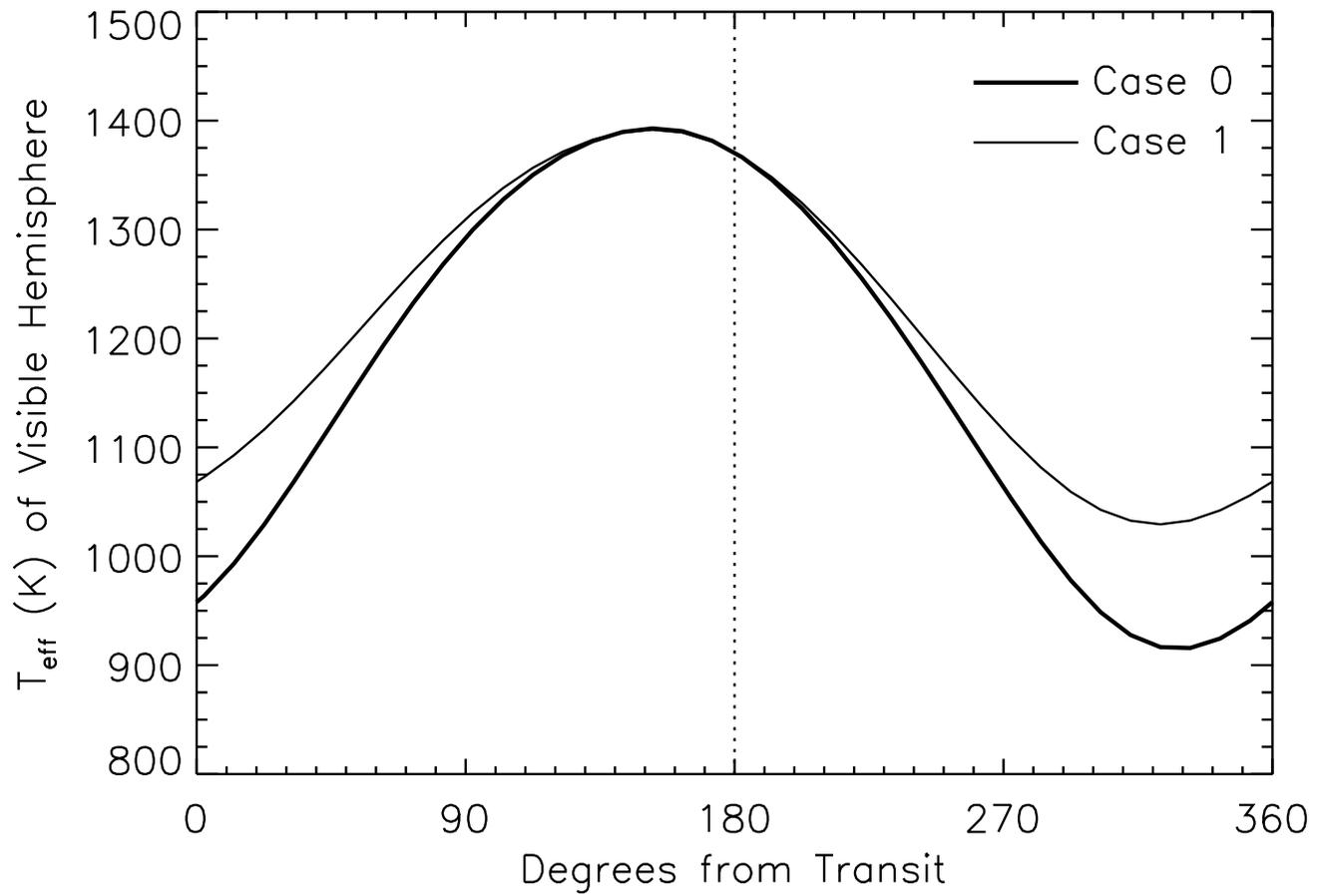}
\caption{Effective temperature of the visible hemisphere of \hd~as a function
of orbital phase for two chemistry cases.  
\label{figure:Teff}}
\end{figure}

One can integrate the spectrum of the planet's visible hemisphere over all
wavelengths, as a function of orbital phase, to determine the apparent
luminosity of the planet at all phases.  Here we divide out $4\pi R^2 \sigma$
to calculate the apparent effective temperature ($T_{\rm eff}$) of the visible
hemisphere.  This is plotted in \mbox{Figure~\ref{figure:Teff}} for Cases 0 and 1.  We can
clearly see that in both cases, the time of maximum flux precedes the time of
secondary eclipse by $\sim$27 degrees, or 6.3 hours.  Since the spectra of the
two cases overlap around the time of secondary eclipse, so do the plots of
$T_{\rm eff}$.  At other orbital phases, the Case 0 curve always plots lower.
The largest effect is before the time of transit.  The effect is tied to the
CH$_4$ abundance.  If CH$_4$ is able to attain a large mixing ratio, it leads
to an atmosphere that has a higher opacity, meaning one can not see as deeply
into the atmosphere.  One then reaches an optical depth of $\sim$1 higher in the
atmosphere, which is significantly cooler for night-side \emph{P--T} profiles,
leading to a lower $T_{\rm eff}$.  The point of minimum planetary flux
precedes the transit by 31 to 37 degrees, depending on the chemistry.

In their (very similar) previous dynamical model, \citet{CS05} predicted a
time of maximum planetary flux of 60 degrees (or 14 hours) before secondary
eclipse.  The large timing difference between that work and this one is due
almost entirely to the choice of ``photospheric pressure" made in
\citet{CS05}.  They chose a pressure of 220 mbar in their work, which is
deeper than the ``mean" photosphere that we find here.  At higher pressure, the
radiative timescales are longer, such that winds are able to blow the
atmosphere's hottest point farther downstream. \citet{CS05} and \citet{Showman06} previously discussed how their prediction varied as a function of the chosen photospheric pressure.

\section{Infrared Light Curves}
\subsection{Spitzer Bands}

\begin{figure}
\plotone{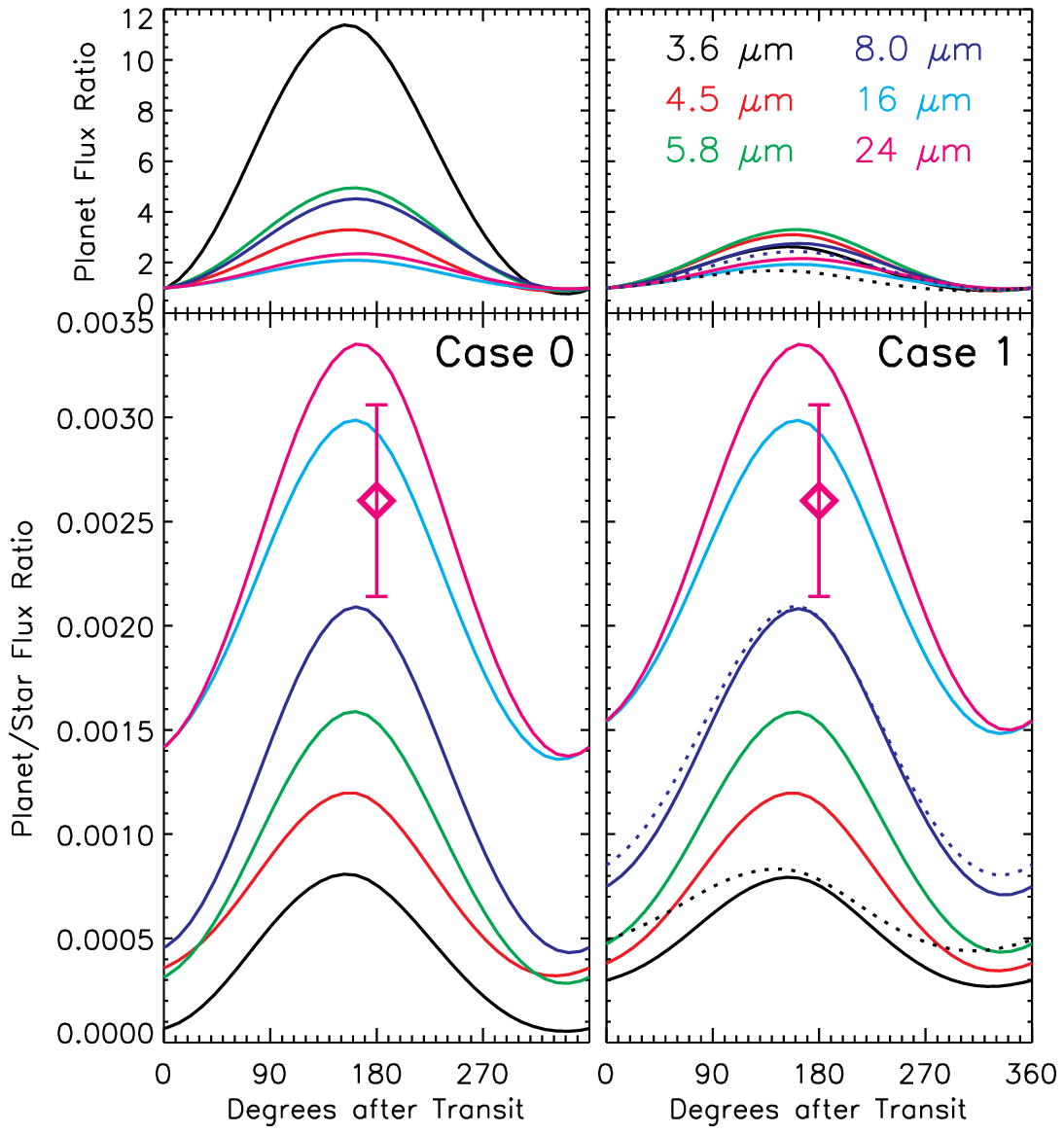}
\caption{Planetary flux in \emph{Spitzer} bands as a function of planetary
orbital phase.  The left panels are Case 0 and right panels Case 1.  The
bottom panels show the \hd~planet-to-star flux ratios.  The top panels show
the planetary flux at every phase divided by the planetary flux seen during
the transit (at zero degrees).  The dotted lines indicate predictions for Case
2 chemistry.  At 180 orbital degrees is the \citet{Deming05b} secondary
eclipse observation at 24 $\mu$m, with one-sigma error bars
(0.00260$\pm$0.00046).  This is shown in magenta, which is the color used for
the 24 $\mu$m curve.
\label{figure:spitzer}}
\end{figure}

For the foreseeable future, the most precise data for understanding the
atmospheres of hot Jupiters will come from the \emph{Spitzer Space Telescope}.
For \hd, only an observation at 24 $\mu$m (the shortest wavelength MIPS band),
has been published.  It seems likely that observations in all four
IRAC bands (3.6, 4.5, 5.8, and 8.0 $\mu$m), as well as IRS at 16 $\mu$m, will be
obtained within the next year or two.  As such, we have integrated the
spectra of our planet models and a \citet{Kurucz93} model of star HD 209458
over the \emph{Spitzer} bands in order to generate planet-to-star flux ratios
as a function of orbital phase.  These are plotted for Cases 0 and 1 in
\mbox{Figure~\ref{figure:spitzer}}.  The stellar model fits the stellar
parameters derived in \citet{Brownetal01} and is the same model used in
\citet{Fortney05}.

The behavior of the planet-to-star flux ratios is quite interesting.  While
the $T_{\rm eff}$ of the planet was found to reach a maximum 27 degrees before
secondary eclipse, the behavior in individual bands is more varied.  For
instance, in both chemistry cases, the planetary flux in the 24 $\mu$m band
peaks only 15 degrees before secondary eclipse.  This is because the
``photospheric pressure'' is at a lower pressure in this band than the
planet's ``mean photospheric pressure.''  As previously discussed, at lower
pressures, the radiative time constants are shorter, and the atmosphere is
able to more quickly adjust to changes in incident flux.  At higher pressure,
winds are better able to blow the planet's hot spot downstream.  One should
keep in mind that the light curves generated are a function of the dynamical
calculation and the radiative transfer.  It is the radiative transfer
calculation that determines how deeply into the atmosphere (and therefore, to
what temperature) we see.

The 3.6 $\mu$m band peaks earliest, $\sim$27
degrees before transit, as the $T_{\rm eff}$ does as well.  This band shows a
15-fold variation in flux (peak to trough) as a function of orbital phase for
Case 0 because it encompasses a significant CH$_4$ absorption feature that
waxes and wanes.  Since the abundances of CH$_4$ and CO are not free to vary
in Case 1, the flux ratios in this case do not show the large amplitudes found
in some bands in Case 0.  The dotted lines in the Case 1 panels are for Case 2
in the 3.6 and 8.0 $\mu$m bands, where CH$_4$ absorption occurs.  The flux
variation in these bands is even further reduced as CH$_4$ absorption is not
seen.  (See \mbox{Figure~\ref{figure:cases}}.)  At 24 $\mu$m, our model is 1.6
sigma higher than the secondary eclipse data point published by
\citet{Deming05b}, indicating that the planet may be dimmer at 180 orbital
degrees than we predict.  In all bands, differences in chemistry between the
two cases have essentially no effect on the timing of the maxima in planetary
flux.  However, since the night side is much more sensitive to chemistry, the
minima can vary by as much as 20 degrees in bands that are sensitive to CH$_4$
absorption.  

\subsection{Standard Infrared Bands}

The results of ground-based observations of flux from hot Jupiters
have been mixed.  All searches for visible light have yielded only upper
limits \citep{Charb99, Collier02, Leigh03}, which have ruled out some models
with bright reflecting clouds.  In the near infrared, specifically for \hd,
\citet{Rich03,Rich03b} have constrained molecular bands of CH$_4$ and H$_2$O.
The constraint on emission at 2.3 $\mu$m between H$_2$O absorption features
was shown in \mbox{Figure~\ref{figure:cases}}.  The predicted planet-to-star
flux ratio really does not become favorable until wavelengths longer than 3
$\mu$m, which may continue to challenge observers.

\begin{figure}
\plotone{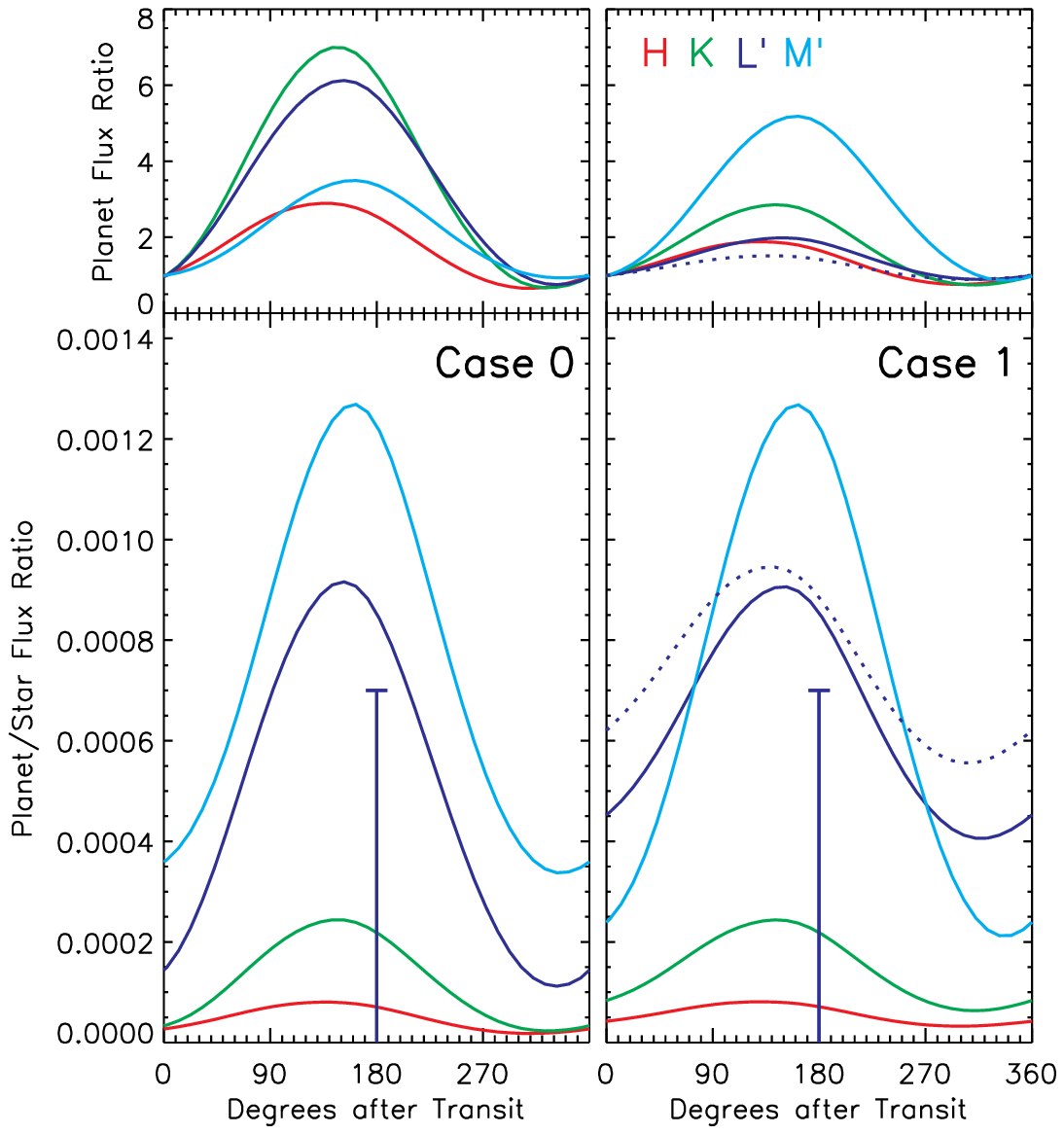}
\caption{Planetary flux in MKO infrared bands as a function of planetary
orbital phase.  The left panels are Case 0 and right panels Case 1.  The
bottom panels show the \hd~planet-to-star flux ratios.  The top panels show
the planetary flux at every phase divided by the planetary flux seen during
the transit (at zero degrees).  The dotted line indicates a prediction for
Case 2 chemistry.  At 180 orbital degrees is the \citet{Deming05c} secondary
eclipse observation at L$^{\prime}$ band, with one-sigma error bars.  The
observation is -0.0007$\pm$0.0014, so the actual data point is well off of the
bottom of the plot.  This is shown in dark blue, which is the color used for
the L$^{\prime}$ curve.
\label{figure:mko}}
\end{figure}

In \mbox{Figure~\ref{figure:mko}}, we plot planet-to-star flux ratios in the
Mauna Kea Observatory (MKO) H ($\sim$1.6 $\mu$m), K ($\sim$2.2 $\mu$m),
L$^{\prime}$ ($\sim$3.8 $\mu$m), and M$^{\prime}$ ($\sim$4.7 $\mu$m) bands.
Since the wavelength ranges of the L$^{\prime}$ and M$^{\prime}$ bands have
significant overlap with the IRAC 3.6 and 4.5 $\mu$m bands, the predicted
ratios are quite similar.  In dark blue, we show the L$^{\prime}$ band upper
limit of -0.0007$\pm$0.0014 from \citet{Deming05c}.  The model is just above the 1
sigma error bar.\footnote{As described in \citet{Deming05c}, this observation
was actually performed in a narrow band centered on 3.8 $\mu$m, within the
standard L$^{\prime}$ band.  Our calculated planet-to-star flux ratio at
secondary eclipse increases by 4\% when using this narrow band, compared to
standard L$^{\prime}$.  Since this is a small correction, our conclusions are
unchanged.}  This is a better match than is attained with one
dimensional radiative equilibrium models \citep{Deming05c}, which predict a flux peak just short of 4 $\mu$m, as shown in
\mbox{Figure~\ref{figure:cases}}.  The \citet{Fortney05} \hd~radiative equilibrium model, which has somewhat muted flux peaks compared with similar models by other authors, has a planet-to-star flux ratio of 0.00114 in L$^{\prime}$ band. In dotted blue in the Case 1 panels is our
L$^{\prime}$ band (which encompasses CH$_4$ absorption) prediction for Case 2
chemistry.  Looking at shorter wavelengths, in the H and K bands, the planet is
predicted to be dimmer, while the star is brighter, leading to low flux
ratios.  The peak emission in these bands does occur earlier than in the
\emph{Spitzer} bands.  For instance, the H band peak is 42 orbital degrees
before secondary eclipse.

\section{Discussion}
\subsection{Effective Temperature}
For the CS06 dynamical model of the atmosphere of \hd, we find that the apparent $T_{\rm
eff}$ of the visible hemisphere is strongly variable, with a maximum of 1390 K
and a minimum of 915 K for equilibrium chemistry and 1025 K for (the probably
more realistic case of) disequilibrium chemistry.  This leads to a luminosity
of the visible planetary hemisphere that varies by factors of 5.3 and 3.4,
respectively for these two cases.  For the one dimensional \citet{Iro05} planet-wide  profile, we derive a $T_{\rm eff} = 1325$ K, 1\% less than found by the authors.  The day-side $T_{\rm
eff}$ for the dynamical model is not significantly larger than this planet-wide $T_{\rm eff}$.  As an additional comparison, we calculate the mean of the planetary luminosity over the entire orbit and then convert to $T_{\rm eff}$ to find a mean $T_{\rm eff}$ for Cases 0 and 1.  This gives $T_{\rm eff}$s of 1195 K and 1227 K, respectively, showing that the CS06 model, at the pressures levels that radiate to space, is as a whole somewhat colder than the one dimensial profile of \citet{Iro05}.  This is likely a consequence of the radiative forcing scheme employed in CS06, which will be reinvestigated when models that consistently couple radiative transfer and dynamics are developed.

The $\sim$400 K $T_{\rm eff}$ contrasts found are
predominantly due to changes in the temperature structure of the hemisphere
that is visible as function of orbital phase.  In addition, atmospheric
opacity makes an important contribution when the CH$_4$/CO ratio is free to
vary with pressure and temperature.  When the CH$_4$ mixing ratio is below that predicted by
equilibrium chemistry, this leads to a lower opacity atmosphere, for a given
\emph{P--T} profile.  While the few individual bands of CO are somewhat
stronger than those of CH$_4$, CH$_4$ absorption across the planet's broad the
2-10 $\mu$m flux peak dominates over the two bands of CO.  One is able to see
more deeply into a CH$_4$-depleted atmosphere, leading to a higher $T_{\rm eff}$.

\subsection{\hd~Infrared Data}
Currently, the only published secondary eclipse datum from \emph{Spitzer} for \hd~is the 24
$\mu$m detection of \citet{Deming05b}.  The models presented here have a
planet-to-star flux ratio during secondary eclipse that is 1.6 sigma larger
than this observational data point.  Together with our excellent agreement
with the ground-based data at 2.3 $\mu$m from \citet{Rich03b} and
\citet{Seager05}, and our 1.1 sigma difference with the 3.8 $\mu$m data from
\citet{Deming05c}, we regard this as excellent agreement---significantly
better than has been previously obtained with one dimensional radiative equilibrium models.

It is
interesting to discuss a few issues that arise if the flux ratios are indeed
10-25\% less than we calculate here, as indicated by the L$^{\prime}$ and 24 $\mu$m band data.  For instance, perhaps day-night
temperature contrasts in the atmosphere are not as large as predicted by CS06,
leading to smaller deviations from a ``planet-wide'' $T_{\rm eff} \sim 1200$K.
This may involve radiative time constants that are significantly
longer than predicted by \citet{Iro05}, winds faster than predicted by CS06,
or both.  It would be worthwhile for other groups that possess hot Jupiter
radiative transfer codes to compute radiative time constants at these
temperatures and pressures.  This is an area that we will pursue in the near
future.

Another possibility for a smaller planet-to-star flux ratio during secondary
eclipse would be if the planet had a larger Bond albedo than calculations
currently indicate.  This would mean less absorbed stellar flux and
correspondingly lower temperatures everywhere on the planet.  Cooler temperatures everywhere on the planet would lead to chemical abundances that differ from our previous cases.  More CH$_4$ would form, at the expense of CO, which would also lead to a slightly higher mixing ratio for H$_2$O, which shares oxygen with CO.

Hot Jupiters are believed to have very low Bond albedos--on the order of 90\% or more of incident stellar flux is expected to be absorbed.  In \citet{Fortney05}, we found that our one-dimensional model atmosphere for
\hd~scattered only 8\% of incident flux.  For \T, this was 6\%.  To date,
there is at least a hint that the Bond albedo of \T~may have been
underestimated.  \citet{Charb05}, under the assumption that the planet emits
as a blackbody, determined a Bond albedo of $0.31 \pm 0.14$ from their
\emph{Spitzer} IRAC observations at 4.5 and 8.0 $\mu$m.  Perhaps hot Jupiters are not quite as hot as had been previously thought.  However, this
determination should be considered very preliminary at this time.

\citet{Richardson06} have recently observed the \emph{transit} of 
\hd~at 24 $\mu$m, as well.  From an observed transit depth of $0.0149 \pm 
0.0003$, they determined the radius of the planet in this band to be 1.26 
$\pm 0.08$ \rj, which includes uncertainties in the stellar radius.  Our 
model predicts a change in the apparent planet-to-star flux ratio of 
$\sim$0.00008 during the 20 orbital degrees of the \citet{Richardson06} 
transit observations, $\sim$4 times smaller than their uncertainty, and 
hence 
too small to have been detected.  As was seen in \mbox{Figures~\ref{figure:spitzer}} and \ref{figure:mko}, the change in planetary flux as a function of orbital phase is not as pronounced near the time of transit (and secondary eclipse) as it is at other phases.

\subsection{Temperature Sensitivity}
\begin{figure}
\plotone{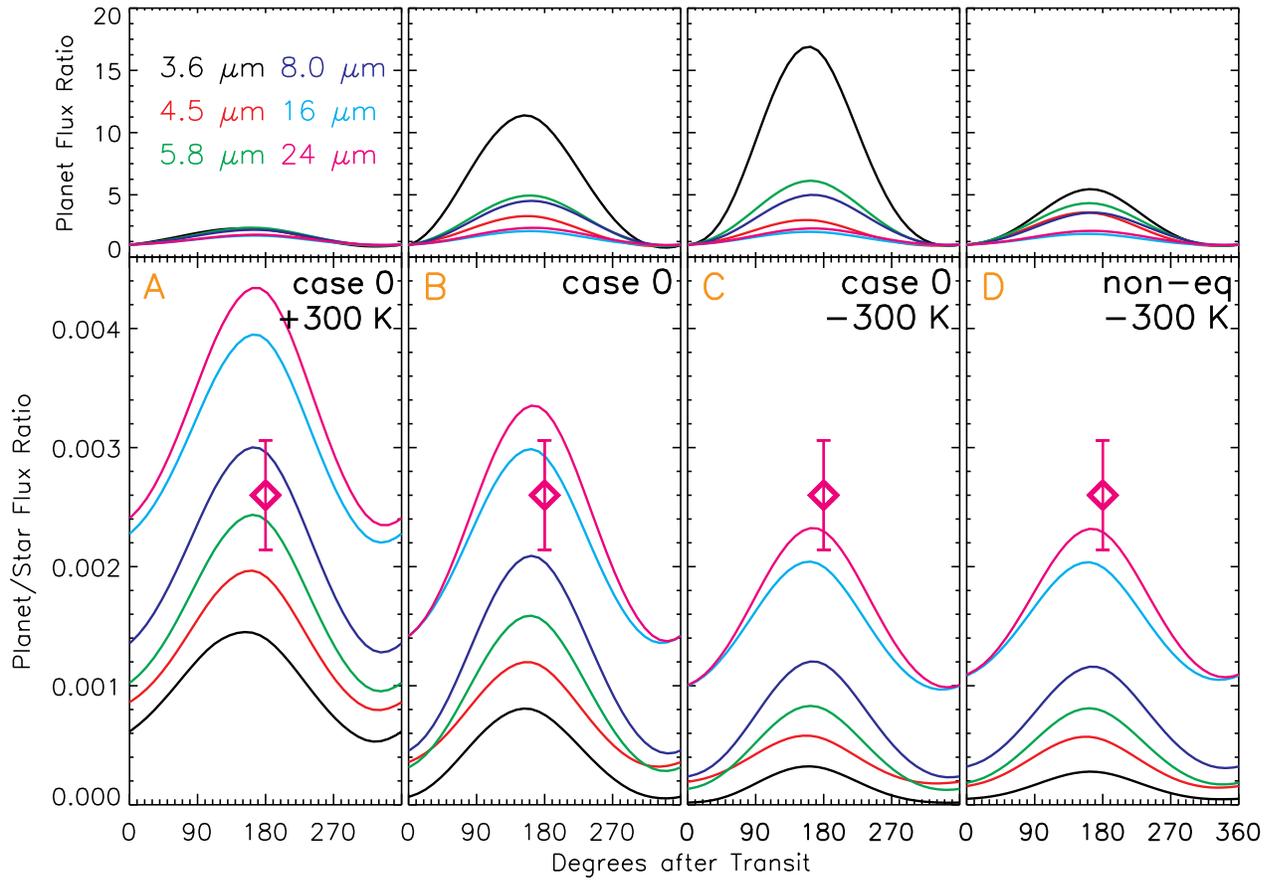}
\caption{Planetary flux in \emph{Spitzer} bands as a function of planetary
orbital phase.  The lightcurves in panels A, B, and C assume equilibrium chemistry while panel D uses non-equilibrium chemistry.  The
bottom panels show the \hd~planet-to-star flux ratios.  The top panels show
the planetary flux at every phase divided by the planetary flux seen during
the transit (at zero degrees).  The left panel (A) shows the CS06 simulation with a 300 K increase in temperature.  Panel B shows the nominal Case 0 simulation, previously shown in \mbox{Figure~\ref{figure:spitzer}}a.  Panel C shows the CS06 dynamical simulation with a 300 K temperature decrease.  The right-most panel (D) is also the simulation with the 300 K temperature decrease, but uses non-equilibrium chemistry with a fixed CH$_4$/CO ratio of 0.20. (See text.)  For all panels, the \citet{Deming05b} datum at 24 $\mu$m is shown.
\label{figure:t3}}
\end{figure}

To illustrate the sensitivity of our results to temperature 
changes, we have computed spectra as a function of orbital phase, using 
equilibrium (Case 0) chemistry for two additional models.  These are additional dynamical models described in CS06, in which the base \emph{P--T} profile of \citet{Iro05} has been increased or decreased by 300 K, with correspondingly warmer or colder night sides.  The full dynamical simulations have been run again with these parameters.  The resulting light curves, in \emph{Spitzer} bands, are shown in \mbox{Figures~\ref{figure:t3}}a, and c.  For the ``cold'' (-300 K) model CH$_4$ is the dominant carbon carrier on the night hemisphere, and CO on the day hemisphere, leading to large flux variation in excess of that found for our nominal CS06 simulation, as shown in \mbox{Figure~\ref{figure:t3}}.  For the ``hot'' (+300 K) model, on both the day and night hemispheres, CO is the dominant carbon carrier.  Methane absorption features are very weak on the night side, leading to flux variations in every band no larger than a factor of 2.7 from peak to trough.  This model is somewhat similar to our earlier Case 2, but at warmer temperatures, as at all phases CO is the dominant carbon carrier.  Fluxes are everywhere greater in the hot model than the nominal model, and everywhere less in the cold model than in the nominal model.  This is is due to the differing atmospheric temperatures.  The cold model best fits the \citet{Deming05b} datum at 24 $\mu$m.  In addition, the phase of maximum and minimum flux in a given band can change by up to $\sim$10 orbital degrees between these simulations, due both to differing chemistry and atmospheric dynamics.  As discussed in \citet{CS05} and CS06, while these models all possess similarly strong east-to-west jets, the dynamical atmospheres differ slightly in detail.

CS06 also examined non-equilibrium CH$_4$/CO chemistry for these models.  For the ``cold'' (-300 K) model, non-equilibrium chemistry was important, and the CH$_4$/CO ratio at $P < 1$ bar became homogenized at 0.20 around the entire planet.  This ratio was 0.014 for the nominal Case 1 decribed earlier.  For the cold model equilibrium chemistry would predict a night side dominated by CH$_4$ and a day side dominated by CO.  For the ``hot'' model (+300 K) both equilibrium and non-equilibrium chemistry predicts that CO is the dominant carbon carrier on both the day and side hemispheres.  \mbox{Figure~\ref{figure:t3}}d shows our computed flux ratios for  non-equilibrium chemistry for the cold case.  As was shown previously for our nominal case, flux variation is smaller with non-equilibrium chemistry, because the mixing ratios of CH$_4$, CO, and H$_2$O are the same on the night and day hemispheres.  Again, because of the nearly isothermal temperature structure of the day-side, chemical abundances have little effect on the day-side planet-to-star flux ratios, which are nearly equal for these two chemistry cases.

These additional cases further highlight the importance of CH$_4$/CO chemistry in the computation of infrared light curves.  Infrared fluxes as a function of orbital phase are sensitive to the temperatures of the hemisphere facing the observer, as well as the abundances of important molecular absorbers.  For planetary \emph{P--T} profiles that cross important CH$_4$/CO chemical boundaries, as most hot Jupiters surely do, whether or not these species are found in their equilibrium mixing ratios has a major impact on resulting infrared flux in \emph{Spitzer} bands, especially on the night side.

\subsection{The Future \& Conclusions}
From the size of the error bar from the \citet{Deming05b} 24 $\mu$m
observations, it is clear that, could this instrument stability be sustained over the
course of tens of hours of observations, the change in flux over time
presented here could be detected.  One-half of an orbital period for \hd~is 42
hours.  If flux in the 8.0 $\mu$m band for \hd~is higher than models predict,
as was the case for \T~\citep{Fortney05, Barman05}, then this would be an
attractive band as well, as the error bars should be smaller.

What might one
hope to see with sustained observations?  If the peak in infrared flux does
occur only $\sim$25-30 orbital degrees before secondary eclipse, this would be
difficulty to discern.  However, the detection of any sort of ramp up in flux
from the time of transit to secondary eclipse would give us important
information on the day-night temperature contrast.  The recently discovered
transiting planet \he~\citep{Bouchy05} would likely be an even more attractive
target, as the planet-star flux ratios are likely twice as large
\citep{Fortney06, Deming06}, and the orbital period is $\sim$40\% shorter.  We
predict secondary eclipse planet-to-star flux ratios for this system in
\citet{Fortney06}; our calculation at 16 $\mu$m is a good match to the
published observation of \citet{Deming06}.

A clear prediction from our calculations here is that when one uses realistic
non-equilibrium chemistry calculations, the change in planetary flux as a
function of orbital phase is greatly reduced, relative to equilibrium
chemistry calculations, because the atmosphere's composition is fixed.  For the \emph{Spitzer} bands, for Case 0, the maximum
flux variation is in the 3.6 $\mu$m band, which varies by a factor of 15 from
peak to trough.  The minimum variation is a factor of 2.2, in the 16 $\mu$m
band.  For Case 1, this variation drops significantly, and the maximum
variation is a factor of 3.7, in the 5.8 $\mu$m band, and the minimum is 2.0,
again in the 16 $\mu$m band.  We suggest that the 5.8 and 8.0 $\mu$m bands may
be the best \emph{Spitzer} bands in which to search for flux variations as a
function of orbital phase, as these bands combine a high planet-to-star flux
ratio and the sensitivity and stability of the IRAC detectors.

If the day-side thermal emission of hot Jupiters is similar to a blackbody, as
we find here, problems arise with the notion that the emission can be used to
characterize the atmospheric chemistry from secondary eclipse observations.  Absorption features due to CH$_4$, CO, and H$_2$O would be nonexistent or extremely weak.
Observations at other orbital phases would then take on additional importance.

Given the significant spectral differences between our model and radiative-equilibrium models, it is clear that more work in this area is certainly warranted.  We note that blackbody-like hot Jupiter emission can simultaneously explain all observations to date.  First is the secondary eclipse observation at 24
$\mu$m by \citet{Deming05b}, a clear detection.  Second is the very low flux
ratio upper limit in L$^{\prime}$ band by \citet{Deming05c}.  Third is the 2.3
$\mu$m relative flux observation of \citet{Rich03b}, which most
one-dimensional models cannot fit \citep{Seager05}.  Fourth is the set of
observations of \T~by \citet{Charb05}, who found an infrared spectral slope
from 4.5 to 8 $\mu$m that was redder than that found by most one-dimensional
models \citep{Fortney05, Barman05}.  Additional observations of these planets,
especially in the \emph{Spitzer} 3.6 $\mu$m band, which catches much of the
predicted 4 $\mu$m flux peak, would strengthen or refute this argument and
provide a critical test for the CS06 dynamical simulation.  \citet{Deming05c}
posited, perhaps with a wink, that a blackbody-like spectrum would
be more consistent with observations to date than any published hot Jupiter model.  Due to
the dynamically altered temperature structure of the atmosphere of \hd, we find that this could be the reality.
\\
\\
We acknowledge support from NASA Postdoctoral Program (NPP) and Spitzer Space Telescope fellowships
(J.~J.~F.), NASA GSRP fellowship NGT5-50462 (C.~S.~C.), NSF grant AST-0307664 (A.~P.~S.), and NASA grants NAG2-6007 and NAG5-8919 (M.~S.~M.).


\begin{deluxetable}{ccccc}
\center
\tablecolumns{5}
\tablewidth{0pc}
\tablecaption{Chemistry Cases}
\tablehead{
\colhead{Case \#} & \colhead{X$_{\rm CH_4}$} & \colhead{X$_{\rm CO}$} & \colhead{X$_{\rm H_2O}$} & \colhead{CH$_4$/CO ratio}}
\startdata 
0 & equil. & equil. & equil. & equil.\\ 
1 & 6.60E-6 & 4.81E-4 & 2.75E-4 & 0.014\\ 
2 & 1.50E-8 & 4.88E-4 & 2.69E-4 & 3.0E-5\\ 
\enddata
\tablecomments{Mixing ratios are given for each chemistry case.  Abundances from equilibrium chemistry calculations are used when ``equil.'' is specified.}
\end{deluxetable}

\end{document}